\documentclass[aps, prl, two column]{revtex4}   
\usepackage[british]{babel}
\usepackage{latexsym}
\usepackage{graphicx}

\begin{document}

\title{Conformal symmetry of brane world effective actions}
\author{Paul L. McFadden}
\email{p.l.mcfadden@damtp.cam.ac.uk}
\author{Neil Turok}
\email{n.g.turok@damtp.cam.ac.uk}
\affiliation{DAMTP, CMS, Wilberforce Road, 
  Cambridge, CB3 0WA, UK}
\affiliation{African Institute for Mathematical Sciences, 6 Melrose
  Road, Muizenberg, Cape Town 7945, South Africa} 
\date{\today}

\newcommand{\nc}{\newcommand}
\nc{\rnc}{\renewcommand}
\nc{\eg}{\textit{e.g. }}
\nc{\dx}{\mathrm{d} ^4 x}
\nc{\D}{\partial}
\rnc{\d}{\mathrm{d}}
\nc{\tr}{\mathrm{Tr}}
\nc{\gdxdx}{ g_{\mu \nu}(x) \mathrm{d}x^\mu \mathrm{d}x^\nu}
\nc{\ndxdx}{ \eta _{\mu \nu} \mathrm{d}x^\mu \mathrm{d}x^\nu}
\nc{\tgdxdx}{\tilde{g}_{\mu \nu}(x) \mathrm{d}x^\mu \mathrm{d}x^\nu}
\nc{\dxdx}{\mathrm{d}x^\mu \mathrm{d}x^\nu}
\nc{\tz}{\tilde{z}}
\nc{\g}{g_{\mu \nu}}
\nc{\gpm}{g^\pm _{\mu \nu}}
\nc{\gp}{g^+ _{\mu \nu}}
\nc{\gm}{g^- _{\mu \nu}}
\nc{\tg}{\tilde{g}_{\mu \nu}}
\rnc{\[}{\begin{equation}}
\rnc{\]}{\end{equation}}
\nc{\bea}{\begin{eqnarray}}
\nc{\eea}{\end{eqnarray}}
\nc{\ie}{\textit{i.e. }}
\nc{\dw}{\mathrm{d}\Omega _2^2}
\rnc{\tt}{\rightarrow} 
\rnc{\inf}{\infty}
\rnc{\l}{L}

\begin{abstract}
A simple derivation of the low-energy effective action for brane
worlds is given, highlighting the role of conformal invariance.  
We show how to improve the effective action for  
a positive- and negative-tension brane pair
using the AdS/CFT correspondence.  

\end{abstract}

\maketitle

One of the most striking ideas to emerge from string theory is that
the universe we inhabit may be a brane embedded in, or bounding, a
higher-dimensional spacetime.  The brane construction naturally removes the 
extra dimensions from view, and gives a different perspective on the 
nature of the gravitational force.  It
also leads to important restrictions on the 
form of the low-energy four-dimensional effective action.

In this article, we show in particular how the brane
construction automatically implies conformal invariance
of the four-dimensional effective theory.
This explains the detailed form of the low-energy effective action, 
previously found using other methods.  
The AdS/CFT correspondence may then be used to improve
the effective description, and we show how this works in detail for
a positive- and negative-tension 
brane pair.

We start by considering a pair of four-dimensional 
positive- and negative-tension
$Z_2$-branes bounding a five-dimensional bulk with a negative
cosmological constant \cite{RS1}. This is the 
simplest setting incorporating branes with a non-trivial
warp factor in the bulk.  
As is well known, the model possesses a
one-parameter family of static solutions representing flat branes
located at arbitrary $Y$ in a static AdS bulk $\d Y^2 +
e^{2Y/\l}\ndxdx$, where $L$ is the AdS radius, 
$x^\mu$, $\mu=0,1,2,3$, parameterize the four dimensions tangent to the
branes and 
$Y$ parameterizes the dimension normal to the branes.
The locations of the branes, $Y^\pm$, are moduli.

For the general, non-static solution to the same model it is
convenient to choose coordinates in which the bulk metric takes the
form
\[
\label{bulk_metric}
\d s^2 = \d Y^2 + g_{\mu\nu}(x,Y)\dxdx .
\]
The brane loci are now $Y^\pm (x)$ and the metric induced on each brane is 
\[
\label{brane_metric}
\gpm (x) = \D _\mu Y^\pm (x) \D _\nu Y^\pm (x) + \g (x,Y^\pm (x)).
\]

At low energies we expect the configuration to be completely
determined by the metric on one brane and the normal distance to the
other brane, $Y^+ - Y^-$.  That is, we are looking for a
four-dimensional effective theory consisting of gravity plus one
physical scalar degree of freedom.  What we will now show is that this
theory may be determined on symmetry grounds alone.  (See also
 \cite{K&S} for related ideas).

The full five-dimensional theory is diffeomorphism
invariant.  This invariance includes the special set
of transformations
\[
\label{condn1}
Y' = Y + \xi ^5(x), \ \ \ x'^\mu = x^\mu + \xi^\mu (x,Y),
\]
with $\xi^\mu (x,Y)$ satisfying 
\[
\label{condn}
\D_Y \xi ^\mu (x,Y) = -g^{\mu\nu}(x,Y)\D _\nu\xi ^5 (x),
\]
which preserve the form (\ref{bulk_metric}) of the metric.  
Equation (\ref{condn}) may be integrated to give $
\xi^\mu(x,Y) = \xi^\mu(x,Y^-(x))
- \int_{{Y^-(x)}}^Y dY g^{\mu\nu}(x,Y) \partial_\nu \xi^5 (x)$, where
$\xi^\mu(x,Y^-(x))$ are the parameters of a four-dimensional diffeomorphism
on the minus brane.
The
transformation (\ref{condn1}) displaces the $Y^\pm (x)$ coordinates of
the branes, $Y^\pm (x) \tt Y^\pm (x) + \xi ^5-\xi^\sigma \D _\sigma
Y^\pm (x)$,
and alters $g^{\mu\nu}(x,Y)$ via the usual Lie derivative.  Using
(\ref{condn}), one finds that the combined effect on each brane metric 
(\ref{brane_metric}) is the four-dimensional diffeomorphism $x'^\mu =
x^\mu + \xi^\mu (x, Y^\pm (x))$.  In fact, by departing from the gauge
(\ref{bulk_metric}) away from the branes, we can construct a five dimensional
diffeomorphism for which 
$\xi^\mu$
vanishes on the branes.  To see this, we can set $\xi^\mu (x,Y)=\xi^\mu _0
(x,Y)-f(Y)\xi^\mu _0 (x,Y^+)$, where $\xi^\mu _0 (x,Y)$ is the
solution to (\ref{condn}) which vanishes on the minus brane and $f(Y)$
is a function chosen to satisfy $f(Y^-) = 0$, $f(Y^+) = 1$, and
$f'(Y^-)=f'(Y^+)=0$ for all $x$.

We conclude that the four-dimensional theory, in which $Y^\pm
(x)$ are represented as scalar fields, must possess a local symmetry
$\xi ^5(x)$ acting nontrivially on those fields.  The dimensionless 
exponentials $\psi^\pm (x) \equiv e^{ Y^\pm (x)/\l}$ transform as
conformal scalars: 
$\psi ^\pm (x) \tt e^{\xi ^5 /\l}\psi^\pm (x)$, while the induced
brane metrics $\gpm$ remain invariant.
The only local, polynomial, two-derivative action
possessing such a symmetry involves gravity with two conformally-coupled
scalar fields.  After diagonalizing and rescaling the fields,
this may be expressed as
\[
m^2\int \dx \sqrt{-g} \left( c_+ \psi^+ \Delta \psi^+ + 
c_- \psi^-\Delta  \psi^-\right) ,
\]
where $\Delta \equiv \Box -\frac{1}{6}R$, $c_\pm = \pm 1$, and $m$ is
a constant with dimensions of mass.  
It should be stressed that the metric $\g$ appearing in this
expression is that of the effective theory, which is in general different to the
induced metric on the branes $\gpm$.  
Potential terms are excluded by the fact that
flat branes, with arbitrary constant $\psi^\pm$, are solutions of the
five-dimensional theory, \ie the $\psi^\pm$ are moduli.   

By construction, the theory possesses local conformal invariance under
\[
\label{conf_transf}
\psi^\pm \tt \Omega (x)^{-1}\psi^\pm, \ \ \ \g \tt \Omega (x)^2 \g .
\]
For $c_+=-c_-$, without
loss of generality we can set $c_+=-1$. Provided $(\psi^+)^2 - (\psi^-)^2 > 0$,
we obtain the usual sign for the Einstein term, so there are no ghosts
in the gravitational sector. We can then set 
$\psi^+=A\cosh{\phi/\sqrt{6}}$ and
$\psi^- = -A\sinh{\phi/\sqrt{6}}$. The field $A$ has the wrong sign
kinetic term, but it can be set equal to a constant by a choice
of conformal gauge. Therefore, in this case there are no physical
propagating ghost fields. In contrast, a similar analysis reveals that
when $c_+=c_-$ the theory 
possesses physical ghosts either in the gravitational wave sector
(wrong sign of $R$) or in the scalar sector, no matter how the 
conformal gauge is fixed. 
We conclude that the low-energy
effective action must be 
\[
\label{conf_action}
m^2 \int \dx \sqrt{-g} \left(- \psi^+ \Delta  \psi^+ +
 \psi^-\Delta  \psi^-\right).
\]

We know from the above argument that the brane metrics are conformally-invariant:
from this and from general covariance they must equal $\g$
times homogeneous functions of order two in $\psi^+$ and $\psi^-$.
But in the model under consideration, we have static
solutions $\gpm = e^{2Y^\pm/\l} \eta _{\mu\nu}$ for all $Y^+>Y^-$.  The only
choice consistent with this and with $(\psi^+)^2 - (\psi^-)^2 > 0$ is
\[
\label{g_eqns}
\gpm = \frac{(\psi^\pm)^2}{6}\g ,
\]
which is a conformally-invariant equation. We have introduced the
numerical factor for later convenience.

It is instructive to fix the conformal gauge in several ways.  First,
set $\psi^+=\sqrt{6}$, so that $\g = \gp$ and the metric appearing in
(\ref{conf_action}) is actually the metric on the plus (positive-tension) brane.  The
action (\ref{conf_action}) then consists of Einstein gravity (with
Planck mass $m$) plus a conformally-invariant
scalar field $\psi^-$ which has to be smaller than
$\sqrt{6}$:  
\[
\label{+action}
m^2 \int \dx \sqrt{-g^+}\left( (1-\frac{1}{6}(\psi^-)^2)R^+-(\D
  \psi^-)^2 \right).
\]
Changing variables to $\chi = 1 - (\psi^-)^2/6$ produces the
alternative form \cite{oldKS}
\[
m^2 \int \dx \sqrt{-g^+}\left(\chi R^+ - \frac{3}{2(1-\chi)}(\D \chi)^2
\right).
\]
Conversely, if we set $\psi^-=\sqrt{6}$, then $\g$ is
the metric on the minus (negative-tension) brane and $\psi^+$, which has to be larger
than $\sqrt{6}$, is a conformally-coupled scalar field.  (However, the relative 
sign between the gravitational and kinetic terms in the action is now wrong, 
and so this gauge possesses ghosts).   
If we add matter coupling to the metric on the plus and minus branes, we find
that matter on the minus brane couples in a
  conformally-invariant manner to the plus brane metric and the field
  $\psi^-$, and conversely for matter on the plus brane.
Note that we are not implying conformal invariance of the matter itself:
it is simply that matter coupled to the brane metrics will be trivially
invariant under the transformation (\ref{conf_transf}) as the brane
metrics are themselves invariant.

A third conformal gauge maps the theory to Einstein gravity with a
minimally-coupled scalar field $\phi$, taking the values $-\infty <\phi <0$. 
Starting from
(\ref{conf_action}), we can set $\psi^+=A\cosh{\phi/\sqrt{6}}$ and
$\psi^- = -A\sinh{\phi/\sqrt{6}}$, as noted earlier, to obtain the action
\[
\label{EFaction1}
-m^2\int \dx \sqrt{-g}\left(A\Delta A+\frac{A^2}{6}(\D \phi
 )^2\right).
\]
Now choosing the conformal gauge $A=\sqrt{6}$ we find
\[
\label{EFaction2}
m^2 \int \dx \sqrt{-g}\left(R+\phi\Box\phi\right) ,
\]
\ie gravity plus a minimally-coupled massless scalar.
In this gauge equations
(\ref{g_eqns}) read:
\[
\label{g_eqns2}
\gp = \cosh ^2 (\frac{\phi}{\sqrt{6}})\g, \ \ \ \gm = \sinh ^2
(\frac{\phi}{\sqrt{6}})\g,
\]
in agreement with explicit calculations in the moduli space approach 
 \cite{Ekpyrotic}.  

The present treatment also goes some way towards explaining the
moduli space results.  For example, the fact that the moduli space
metric is flat is seen to be a consequence of conformal invariance.
Specifically, for solutions with cosmological symmetry one can pick a
conformal gauge in which the metric is static.  The scale factors on
the two branes are determined by $\psi^\pm$.  From
(\ref{conf_action}), the moduli space metric is just two-dimensional
Minkowski space.

A couple of results for conformal gravity follow from the above
discussion.  First, in the $\psi^+ = \sqrt{6}$ gauge, we have
$\psi^-=-\sqrt{6}\tanh{(\phi/\sqrt{6})}$. Any solution for a
minimally-coupled scalar $\phi$, with metric $\g$, thus yields a
corresponding solution for a conformally-coupled scalar $\psi^-$, with
$|\psi^-| < \sqrt{6}$ and metric $\gp$ as in (\ref{g_eqns2}), and vice versa.  
Second, in the $\psi^-=\sqrt{6}$ gauge, we have
$\psi^+=-\sqrt{6}\coth{(\phi/\sqrt{6})}$.  Hence we may also obtain a
solution for a conformally-coupled scalar $\psi^+$, with
$|\psi^+|>\sqrt{6}$ and metric $\gm$ given in (\ref{g_eqns2}).  
Thus solutions to conformal scalar gravity
come in pairs: if $\g$ and $\psi$ are a solution, then
$(\psi^2/6)\g$ and $\tilde{\psi}=6/\psi$ is another solution.
In terms of branes, this merely states that if $\gp$ and
$\psi^-$ are known in the gauge $\psi^+=\sqrt{6}$, then it is
possible to reconstruct $\gm$ and $\psi^+$ in the gauge
$\psi^-=\sqrt{6}$.

The argument given above establishing the conformal symmetry 
of the effective action is of a very general nature: 
the only step at which we specialized to
the Randall-Sundrum model was in the identification of the brane
metrics in terms of the effective theory variables (\ref{g_eqns}).
This required only the knowledge of a one-parameter family of solutions. 

To derive the effective theory for other brane models, it is only 
necessary to generalize this last step.
For example, in the case of tensionless branes 
compactified on an $S_1/Z_2$, the bulk warp
is absent and so we know that a family of static solutions is given 
by the ground state of Kaluza-Klein theory (in which all fields are
independent of the extra dimension, and so the additional $Z_2$
orbifolding present in the tensionless brane case is irrelevant).
Ignoring the gauge fields, the Kaluza-Klein ansatz for the five-dimensional 
metric is 
\[
\d s^2 = e^{2\sqrt{2/3}\phi(x)} \d y^2 + e^{-\sqrt{2/3}\phi(x)}\gdxdx ,
\]
where $\phi$ and $\g$ extremize an action identical to (\ref{EFaction2}).
For branes located at constant $y$, the induced metrics are
$e^{-\sqrt{2/3}\phi}\g$, independent of $y$.

Using the effective action in the form (\ref{EFaction1}), conformal
invariance of the induced brane metrics dictates that
\[
\g ^\pm = A^2 f^\pm (\phi) \g ,
\]
for some unknown functions $f^\pm$.  
Upon fixing the conformal gauge to
$A=\sqrt{6}$ one recovers the action (\ref{EFaction2}), which is
just the standard Kaluza-Klein low-energy effective action.
The functions $f^\pm$ are thus both equal to
$\frac{1}{6}e^{-\sqrt{2/3}\phi}$ and we have
\[
\g^\pm = e^{-\sqrt{2/3}\phi} \g = \frac{1}{6} (\psi^+ + \psi^-)^2 \g.
\]
Note that this is consistent with the $\phi \tt -\inf$ limit of the
Randall-Sundrum theory (\ref{g_eqns2}): as the brane separation goes to
zero, the warping of the bulk becomes negligible and the Randall-Sundrum theory
tends to the Kaluza-Klein limit \cite{tolley}.

We now turn to a discussion of 
the general cosmological solutions representing colliding branes.  
We choose a conformal gauge in which
the metric is static, and all the dynamics are contained in
$\psi^\pm$.  For flat, open and closed spacetimes the spatial Ricci scalar
$R=6k$, where $k = 0$,$-1$ and $+1$ respectively.
The action (\ref{conf_action}) yields the equations of motion
\bea
\ddot{\psi}^\pm &=& -k\psi^\pm \\
(\dot{\psi}^+)^2-(\dot{\psi}^-)^2 &=& -k\left(
(\psi^+)^2-(\psi^-)^2\right).
\eea
For $k=0$ we have the solutions
\[
\psi^+ = -At+B, \ \ \ \psi^-=At+B, \ \ \ t<0
\]
representing colliding flat branes.  It is natural to match $\psi^+$ to
$\psi^-$ across the collision, and vice versa, to obtain $\psi^\pm =
\pm At+B$ for $t>0$.  This solution then describes two branes which
collide and pass through each other, with the plus brane continuing to
a minus brane and vice versa \cite{seiberg,tolley}. 

For $k=-1$, we have the three solutions
\bea
\begin{array}{ccc}
\psi^{(1)} = A\sinh{t}; & A\cosh{t}; & Ae^t, \\
\psi^{(2)} = A\sinh{(t-t_0)}; & A\cosh{(t-t_0)}; & Ae^{t-t_0},
\end{array}
\eea
where we set $\psi^+$ equal to the greater, and $\psi^-$
equal to the lesser, of $\psi^{(1)}$ and $\psi^{(2)}$.  For
$k = +1$, we find the bouncing solutions $\psi^{(1)} =
A\sin{t}$, $\psi^{(2)} = A\sin{(t-t_0)}$.  In the absence of matter on
the minus brane, the $\sin$ and $\sinh$ solutions are singular when the minus
brane scale factor $a_-$ vanishes.  However, matter on the minus brane
scaling faster than $a_-^{-4}$, for example scalar kinetic matter,
causes the solution for $\psi^-$ to bounce smoothly 
at positive $a_-$ because $\psi^-$ has a
positive kinetic term. This bounce is perfectly regular.  
However, the ``big
crunch-big bang'' singularity, occurring when the positive- and negative- tension branes collide,
is unavoidable.

The above example illustrates a general feature of the brane pair
effective action. If the positive- and negative- tension brane 
solutions are continued through the collision
without relabeling
(this means that the orientation of the warp must flip) then the 
four-dimensional 
effective action changes sign.  The relabeling restores the conventional
sign. The same phenomenon is seen in string theories obtained by
dimensionally reducing eleven dimensional supergravity, when the 
eleventh dimension collapses and reappears. 
Brane world black hole solutions with intersecting branes are discussed 
in \cite{Us}.


Recently it has been shown that the AdS/CFT correspondence
 \cite{AdS/CFT} provides a powerful approach to the understanding of brane worlds.
For a single positive-tension brane the four-dimensional effective
description comprises simply Einstein 
gravity plus two copies of the dual CFT \cite{deHaro} (as the $Z_2$
symmetry implies there are two copies of the bulk).  
Notable successes of this program include reproducing the
$O(1/r^3)$ corrections to Newton's law on the brane \cite{Duff}, and
reproducing the modified Friedmann equation induced on the
brane \cite{Gubser, Shiromizu&Ida}. 

Consider for simplicity a single positive-tension brane containing only radiation.
Taking the trace of the effective Einstein equations we find 
\[
\label{trace}
-R = 2(8\pi G_4)<T_{CFT}> ,
\]
as the stress tensor of the radiation is traceless.  The trace anomaly
of the dual $\mathcal{N}=4$ $SU(N)$ super-Yang Mills theory must then
be evaluated.  With the help of the AdS/CFT dictionary,
this quantity may be calculated for the case of cosmological symmetry
as shown in \cite{Henningson&Skenderis}, giving 
\[
\label{trace2}
-R=\frac{L^2}{4}\left(R_{\mu\nu}R^{\mu\nu}-\frac{1}{3}R^2\right).
\]
Here, the usual $R^2$ counterterm has been added to the action in
order to eliminate the $\Box R$ term in the trace, thus furnishing second
order equations of motion.  

For a cosmological metric with scale factor $a$ this becomes
\[
\label{hdoteqn}
2(\ddot{a}a+ka^2)=L^2(k+h^2)\dot{h},
\]
where $h \equiv \dot{a}/a$ and the dot denotes differentiation with
respect to conformal time. 
Re-expressing the left-hand side as $h^{-1}\partial
_t(\dot{a}^2+ka^2)$ we can then integrate to obtain
\[
\label{hinteqn}
h^2+k = \frac{1}{a^2}(B-\frac{1}{4}k^2 L^2) + \frac{1}{4}(h^2+ k)^2 {L^2\over a^2},
\]
where $B$ is an integration constant. Now, we can expect to recover 
Einstein gravity on the brane in the limit when $L\rightarrow 0$, with
other physical quantities fixed. We expand all terms in powers of $L$. 
At leading order we must obtain four-dimensional 
Einstein gravity, for which $8\pi G_4 = 8\pi G_5/L$. So we set
$B\sim (8 \pi G_5 \rho_0/3L) +C$,  where $\rho=\rho_0/a^4$ is the  energy
density of conventional radiation, and 
$C$ is a constant independent of $L$ as $L\rightarrow 0$.
From (\ref{hinteqn}) we then obtain the first correction to $h^2+k$,
namely
\[
h^2+k= \frac{8 \pi G_5
  \rho_0}{3La^2} + \frac{C}{a^2}+
  \frac{(8\pi G_5\rho_0)^2}{36a^6} + O(L),
\]
which, thanks to the CFT contribution, now includes 
the well-known dark energy and $\rho^2$ corrections \cite{bin}.

It should come as no surprise that the AdS/CFT correspondence 
only approximates the Randall-Sundrum setup 
up to first nontrivial order in an expansion in $L$. 
The AdS/CFT scenario involves string theory on $AdS_5\times S_5$.  Since 
$\alpha '\sim \ell_s^2 \sim L^2$ at fixed 't Hooft
coupling, and the masses squared of the 
Kaluza-Klein modes on the $S_5$ are of order $1/L^2$,
we expect nontrivial corrections at second order in an expansion in
$L$. Furthermore, 
one can show from the AdS/CFT dictionary 
that in order
for the $\rho^2$ term to dominate in the modified Friedmann equation,
the temperature
of the conventional radiation must be greater than the 
Hagedorn temperature of the string. Clearly, the AdS/CFT correspondence
cannot describe this situation.

We now extend the AdS/CFT approach to the case of a pair of
positive- and negative- tension 
branes
using the ideas developed earlier in this paper.  
The effective action for a single
positive-tension brane is
\[ 
\label{AdS/CFTaction}
\frac{1}{16\pi G_4}\int\dx \sqrt{-g^+}R^+  + 2W_{CFT}[g^+] + S_{m}[g^+],
\]
where $\gp$ is the induced metric on the brane, $S_{m}$ is the brane
matter action, and $W_{CFT}$ is the
CFT effective action (including the appropriate $R^2$
counterterms).   
Substituting now for $\gp$ using (\ref{g_eqns}), the Einstein-Hilbert
term $\sqrt{-g^+}R^+$ becomes $-\sqrt{-g}\psi^+\Delta\psi^+$.
A negative tension brane may then be incorporated as follows: 
\bea
\label{S2}
\frac{1}{16\pi G_4}\int\dx\sqrt{-g}(-\psi^+ \Delta \psi^+ +
\psi^- \Delta \psi^-) + 2W_{CFT}[g^+] &&\nonumber  \\
-2W_{CFT}[g^-]+S_{m}[g^+]+S_{m}[g^-]. \ \ \ \ \ \ \ \ \ \ \  
\eea
The action for the positive- and negative-tension brane pair 
must take this form in order to correctly
reproduce the Friedmann equation for each brane. 
To see this, consider again the conformal gauge in which the effective
theory metric is static and all the dynamics are contained in
$\psi^\pm$, which play the role of the brane scale factors.
Variation with respect to the $\psi^\pm$ yields the
scalar field equations 
\[
(\psi^\pm)^{-3}\Delta \psi^\pm = 2(8\pi G_4)<T_{CFT}^\pm >,
\]
where the trace anomaly must be evaluated on the induced brane metric
$\gpm$ but $\Delta$ is evaluated on the effective metric $\g$.
The left-hand side evaluates to
$-(\psi^\pm)^{-3}(\ddot{\psi}^\pm+k(\psi^\pm)^2)$.  After
identifying $\psi^\pm/\sqrt{6}$ with $a_\pm$ according to (\ref{g_eqns}),
and then dropping the plus or minus label,
we recover equation (\ref{hdoteqn}).  
From the necessity of recovering the Friedmann equation on each brane
we may also deduce that cross-terms in the action between $\psi^+$ and
$\psi^-$ are forbidden. 

The signs of the gravity parts of the 
action are needed to 
achieve consistency with (\ref{conf_action}).
Consequently, the relative sign between the
gravity plus CFT part of the action and that of the matter is reversed for the
minus brane, consistent with the 
modified Friedmann equations \cite{bin},
\[
\label{FRW}
H^2_\pm = \pm \frac{8\pi G_5\rho_\pm}{3\l} + {(8\pi G_5\rho_\pm)^2 \over 36} 
-\frac{k}{a^2}+ \frac{C}{a^4}, 
\]
where plus and minus label the positive- and negative-tension branes,
and $C$ is again a
constant representing the dark radiation.  

To summarize, we have elucidated the origin of conformal symmetry in
brane world effective actions, and shown how this 
determines the effective action to lowest
order.  When combined with the 
the AdS/CFT correspondence, our approach also recovers the first
corrections to the brane Friedmann equations.

{\it Acknowledgements:} We thank PPARC for support. 
NT is supported by the Darley Professorial Fellowship.

\end{document}